\documentclass[12pt]{article}

\usepackage{amssymb}
\usepackage{amsmath}
\usepackage[top=80pt,bottom=85pt,left=85pt,right=85pt]{geometry}
\usepackage{setspace}
\usepackage{sectsty}
\usepackage{braket}
\usepackage{graphicx,color,subfigure}
\usepackage{xcolor}
\usepackage{mathalfa}
\usepackage{cancel}
\usepackage[debug,pageanchor=false]{hyperref}
\definecolor{link}{rgb}{.8,.15,.1}
\usepackage{mdframed}
\usepackage{soul}
\usepackage{multirow}
\usepackage[section]{placeins}
\hypersetup{colorlinks=true,linkcolor=link,citecolor=link,urlcolor=link,linktocpage}

\newcommand{\vol}{\mathrm{vol}}
\newcommand{\nn}{\nonumber}

\newcommand{\f}{\mathfrak{f}}

\newcommand{\USp}{\mathrm{USp}}
\newcommand{\SU}{\mathfrak{su}}
\newcommand{\SO}{\mathfrak{so}}

\newcommand{\LL}{\mathcal{L}}
\newcommand{\dd}{\mathrm{d}}
\newcommand{\dil}{\Lambda}
\newcommand{\nb}{\nabla}

\newcommand{\bR}{R}

\makeatletter
\@addtoreset{equation}{section}
\makeatother

\allowdisplaybreaks

\begin{document}

\begin{titlepage}

\begin{flushright} \small
UUITP-18/18
\end{flushright}

\begin{center}

\vskip .5in 
\noindent

{
\Large \bf{Perturbing AdS$_6 \times_w S^4$$\,$: \\ \vspace{.1cm} linearised equations and spin-2 spectrum}
}

\bigskip\medskip

Achilleas Passias$^1$ and Paul Richmond$^2$\\

\bigskip\medskip

{
\small 
$^1$Department of Physics and Astronomy, Uppsala University,\\
Box 516, SE-75120 Uppsala, Sweden \\	
     \vspace{.3cm}
$^2$Dipartimento di Fisica, Universit\`a di Milano--Bicocca, \\ 
Piazza della Scienza 3, I-20126 Milano, Italy \\ 
and \\ 
INFN, sezione di Milano--Bicocca
}

\vskip .5cm 

{
\small \tt achilleas.passias@physics.uu.se, paul.richmond@mib.infn.it
}

\vskip .9cm 
	     	
{\bf Abstract }

\end{center}

\vskip .1in

\noindent
We initiate the analysis of the Kaluza--Klein mass spectrum of massive IIA
supergravity on the warped AdS$_6 \times_w S^4$ background, by deriving the linearised equations
of motion of bosonic and fermionic fluctuations, and 
determining the mass spectrum of those of spin-2. The spin-2 modes 
are given in terms of hypergeometric functions and a careful analysis of their
boundary conditions uncovers the existence of two branches of mass spectra,
bounded from below. The modes that saturate the bounds belong to short multiplets
which we identify in the representation theory of the $\f(4)$ symmetry
superalgebra of the  AdS$_6 \times_w S^4$ solution.

\noindent
	
\vfill

\eject

\end{titlepage}

\tableofcontents

\section{Introduction}\label{sIntro}

Gauge field theories in five dimensions are non-renormalizable, however string theory predicts the existence 
of strongly coupled superconformal field theories  for certain gauge groups and matter content \cite{Seiberg:1996bd, Morrison:1996xf}. Such an example appears in type I' string theory from a system of D4-branes probing an O8-plane with a stack of D8-branes on top of it. This system admits a supergravity description in massive IIA supergravity \cite{Brandhuber:1999np, Ferrara:1998gv}, and in the near-horizon limit the geometry becomes
a warped product of six-dimensional anti-de Sitter spacetime AdS$_6$, and a four-sphere $S^4$. The warp factor
is singular at the equator of $S^4$, due the presence of the O8-plane, and the internal space is therefore actually a hemisphere. 
The near-horizon background has an exceptional $\f(4)$ symmetry superalgebra, which is the unique superconformal algebra in five dimensions. The bosonic subalgebra of $\f(4)$ is $\SO(2,5) \oplus \SU(2)$, with $\SO(2,5)$ realised as the isometry algebra of AdS$_6$ and the (R-symmetry algebra) $\SU(2)$ as an isometry of $S^4$. The dual superconformal field theory
arises as the strongly-coupled UV fixed point of $\mathcal{N}=1$ supersymmetric $\USp(2N)$ Yang--Mills theory coupled to $N_f < 8$ hypermultiplets in the fundamental representation, and one hypermultiplet in the antisymmetric representation.\footnote{Orbifold generalizations of this system were introduced and studied in \cite{Bergman:2012kr}.}
 
In view of the AdS/CFT correspondence, there is the motivation to study this supergravity background in order 
to learn about the dual field theory. Examples include the calculation of the holographic entanglement entropy
\cite{Jafferis:2012iv}, and the action of probe branes \cite{Bergman:2012qh, Assel:2012nf}. Another such study is that of the Kaluza--Klein mass spectrum, which corresponds to the spectrum of the dual field theory operators. Complete Kaluza--Klein mass spectra have been obtained for anti-de Sitter compactifications whose geometry is a direct product, and the internal space a coset space, typically a sphere; for example \cite{Biran:1983iy, vanNieuwenhuizen:1984iz, Kim:1985ez, Ceresole:1999zs}. In this note we progress towards obtaining the Kaluza--Klein mass spectrum of a warped compactification, by completing the task of obtaining the linearised equations of motion for small fluctuations around the background. Furthermore, we analyse the spectrum of massive spin-2 particles or gravitons in AdS$_6$, uncovering some interesting features\footnote{Spin-2 excitations of flux compactifications with anti-de Sitter, Poincar\'{e} or de Sitter invariance obey the massless scalar wave equation in ten dimensions \cite{Bachas:2011xa}. For other anti-de Sitter backgrounds of massive IIA supergravity, this fact has been exploited in order to study their mass spectrum without deriving the full set of linearised equations of motion \cite{Richard:2014qsa, Pang:2015rwd, Passias:2016fkm, Pang:2017omp}. For AdS$_6$ backgrounds of type IIB supergravity see \cite{Gutperle:2018wuk}.
}. 

The analysis of the spectrum is complicated by the presence of the warp factor as it modifies the differential operators which determine it.
These are differential operators on the internal manifold,
in the present case a four-sphere, and turn out to be warped versions of the Laplace operator
on $S^4$. Hence the standard spherical harmonic analysis is not readily available, a fact that is also
due to the presence of the singularity at the equator. In the case of the spin-2 modes we reduce the problem to solving a 
hypergeometric ordinary differential equation with the mass spectrum determined by imposing appropriate boundary conditions.
Although a rather modest task compared to the analysis of the full spectrum, it already reveals 
some interesting features: we find two branches of spin-2 mass spectra one of which is rather exceptional in that
a certain derivative of the modes is singular. Both branches are bounded from below, and the bound is saturated by modes belonging to short multiplets which we have identified in the work of \cite{Buican:2016hpb, Cordova:2016emh} on representations of the $\f(4)$ superconformal algebra\footnote{See also \cite{DAuria:2000afl}.}. 

The remainder of this note is as follows. In section \ref{sMassIIA} we briefly review massive IIA supergravity and its AdS$_6 \times_ w S^4$ solution. In section \ref{sLinEOM} we present the linearised equations of motion for fluctuations around this solution. In section \ref{sSpin2} we determine the mass spectrum of fluctuations of spin-2. We end in section \ref{sConclusions} with a discussion and comments on future work. Certain technical details are included in appendices.

\section{Massive IIA supergravity and its AdS$_6$ solution}\label{sMassIIA}

In this section we collect the equations of motion of massive IIA supergravity and review the AdS$_6 \times_w S^4$ solution \cite{Brandhuber:1999np}.

\subsection{Equations of motion}\label{ssIIAEOM}

Massive IIA supergravity\footnote{We use the conventions of \cite{Henneaux:2008nr} for the formulation of the theory, with some changes in notation.} in ten dimensions consists of the following bosonic fields: the metric $g$, the dilaton $\Phi$, the field strengths $H_{3}$, $F_{2}$, $F_{4}$, with the subscript denoting their form rank, and the constant ``Romans mass'' $F_0$. The field strengths satisfy the Bianchi identities 
\begin{align}
	\dd H_3  =  0 \, ,\qquad \dd F_2 =  F_0  H_3 \, , \qquad \dd F_4  =  H_3 \wedge F_2 \, ,  \label{Bianchi}
\end{align}
and are given in terms of the potentials $B_2$, $A_1$ and $A_3$ by
\begin{align}
    H_3 &= \dd B_2 \nn \, , \nn \\
	F_2 &= \dd A_1 + F_0 B_2 \, , \nn \\
	F_4 &= \dd A_3 + A_1 \wedge H_3 + \tfrac{1}{2} F_0 B_2 \wedge B_2 \, .
\end{align}
The equations of motion of the bosonic fields are:\footnote{In what follows we will suppress the subscript denoting the rank of a form field whenever its indices appear.}
\begin{align}
	0  &= R_{MN} - \tfrac{1}{2} \partial_M \Phi \partial_N \Phi - \tfrac{1}{16} F_0^2 e^{5\Phi/2} g_{MN} - \tfrac{1}{2} e^{3\Phi/2} \left( F_{MP} F_N{}^P - \tfrac{1}{16} g_{MN} ( F_{2} )^2 \right) \label{IIAEinstein} \\[5pt]
	&- \tfrac{1}{12} e^{\Phi/2} \left( F_{MPQR} F_N{}^{PQR} - \tfrac{3}{32} g_{MN} ( F_{4} )^2 \right) - \tfrac{1}{4} e^{-\Phi} \left( H_{MPQ} H_N{}^{PQ} - \tfrac{1}{12} g_{MN} ( H_{3} )^2 \right) \, , \nn \\[10pt]
	0 &= \nabla^M \nabla_M \Phi - \tfrac{5}{4} F_0^2 e^{5\Phi/2} - \tfrac{3}{8} e^{3\Phi/2} ( F_{2} )^2 - \tfrac{1}{96} e^{\Phi/2} ( F_{4} )^2 + \tfrac{1}{12} e^{-\Phi} ( H_{3} )^2 \, , \label{IIADilaton} \\[10pt]
	0 &= \nabla^M ( e^{-\Phi} H_{MNP} ) - F_0 e^{3\Phi/2} F_{NP} - \tfrac{1}{2} e^{\Phi/2} F_{NPQR} F^{QR} + \tfrac{1}{2\cdot4!4!} \epsilon_{M_1 \cdots M_8 NP} F^{M_1 \cdots M_4} F^{M_5 \cdots M_8} \, , \label{IIA3Form} \\[10pt]
	0 &= \nabla^M ( e^{3\Phi/2} F_{MN} ) + \tfrac{1}{6} e^{\Phi/2} F_{PQRN} H^{PQR} \label{IIA2Form} \, , \\[10pt]
	0 &= \nabla^M ( e^{\Phi/2} F_{MNPQ} ) - \tfrac{1}{144} \epsilon_{M_1 \cdots M_7 NPQ} F^{M_1 \cdots M_4} H^{M_5 \cdots M_7} \, . \label{IIA4Form} 
\end{align}
In the above $R_{MN}$ is the Ricci tensor and $\epsilon_{M_1M_2 \cdots M_{10}}$ the totally antisymmetric tensor. For a $p$-form $\alpha_p$ we have used $( \alpha_p )^2$ to denote the contraction $\alpha_{M_1 \ldots M_p} \alpha^{M_1 \ldots M_p}$. By taking the trace of the Einstein equation \eqref{IIAEinstein} and substituting into \eqref{IIADilaton} we find an alternate equation for the dilaton:
\begin{align}
0 = \nabla^M \nabla_M \Phi - 2 R + g^{MN} \partial_M \Phi \partial_N \Phi + \tfrac{1}{6} e^{-\Phi} ( H_3 )^2 \, , \label{IIADilatonAlt}
\end{align}
where $R$ is the Ricci scalar.

In addition to the bosonic fields, massive IIA supergravity contains the gravitino $\Psi_M$, and the dilatino $\dil$ which are both 32-component Majorana spinors. Their equations of motion are respectively
\begin{align}
	0 &= \Gamma^{MNP} D_{N} \Psi_{P} - \tfrac{1}{4} \dd \Phi \cdot \Gamma^{M} \dil + \tfrac{1}{4} F_0 e^{5\Phi/4} \Gamma^{MN} \Psi_{N} + \tfrac{5}{16} F_0 e^{5\Phi/4} \Gamma^{M} \dil \nn \\[5pt]
	&- \tfrac{1}{8} e^{3\Phi/4} \big( 2 \Gamma^{[M|} F_{2} \cdot \Gamma^{|N]} \Gamma_{11} \Psi_{N} - \tfrac{3}{2} F_{2} \cdot \Gamma^{M} \Gamma_{11} \dil \big) \nn \\[5pt]
	&- \tfrac{1}{8} e^{-\Phi/2} \big( 2 \Gamma^{[M|} H_{3} \cdot \Gamma^{|N]} \Gamma_{11} \Psi_{N} - H_{3} \cdot \Gamma^{M} \Gamma_{11} \dil \big) \nn \\[5pt]	
	&+ \tfrac{1}{8} e^{\Phi/4} \big( 2 \Gamma^{[M|} F_{4} \cdot \Gamma^{|N]} \Psi_{N} + \tfrac{1}{2} F_{4} \cdot \Gamma^{M} \dil \big) \, , \label{IIAGravitino} \\[10pt]
	0 &= \Gamma^M \nabla_M \dil - \tfrac{5}{16} e^{3\Phi/4} F_{2} \cdot \Gamma_{11} \dil + \tfrac{3}{8} e^{3\Phi/4} \Gamma^{M} F_{2} \cdot \Gamma_{11} \Psi_{M} \nn \\[5pt]
	&+ \tfrac{1}{4} e^{-\Phi/2} \Gamma^{M} H_{3} \cdot \Gamma_{11} \Psi_{M} + \tfrac{3}{16} e^{\Phi/4} F_{4} \cdot \dil - \tfrac{1}{8} e^{\Phi/4} \Gamma^{M} F_{4} \cdot \Psi_{M} \nn \\[5pt]
	&- \tfrac{1}{2} \Gamma^{M} \dd \Phi \cdot \Psi_{M} - \tfrac{21}{16} F_0 e^{5\Phi/4} \dil - \tfrac{5}{8} F_0 e^{5\Phi/4} \Gamma^{M} \Psi_{M}  \, , \label{IIADilatino}
\end{align}
where $\nabla_M$ is the usual spin-covariant derivative acting on (vector)-spinors, and $\cdot$ denotes the Clifford product: $\alpha_p \cdot \dil := \frac{1}{p!} \alpha_{M_1 \ldots M_p} \Gamma^{M_1 \ldots M_p} \dil$. The matrices $\Gamma_M$ generate the Clifford algebra ${\rm C}\ell(1,9)$ and satisfy $\{ \Gamma_M, \Gamma_N \}  =2 g_{MN}$. The constant chirality operator is defined as $\Gamma_{11} = \Gamma_0 \ldots \Gamma_9$.

\subsection{The AdS$_6 \times_w S^4$ solution}\label{ssBrandhuberOz}

The AdS$_6 \times_w S^4$ solution of massive IIA supergravity was found \cite{Brandhuber:1999np} by considering the near-horizon limit of a system of D4--D8-branes in the presence of an O8 orientifold plane. In this background (we use $\mathring{}$ above a field to denote its background value) the metric is a warped product of AdS$_6$ and $S^4$ given by\footnote{We work in the Einstein frame. An overall scale
related to the ``trombone symmetry'' of the equations of motion has been set to one. It can be reinstated by $\dd \mathring{s}^2_{10} \to L^2 \dd\mathring{s}^2_{10}$, $\mathring{F}_4 \to L^3 \mathring{F}_4$, $F_0 \to L^{-1} F_0$.}
\begin{align}
	\dd \mathring{s}^2_{10}  =  e^{2A(y)} \left[ \frac{9}{4} \dd s^2_{\mathrm{AdS}_6}(x) + \dd s^2_{S^4}(y) \right] \, . \label{BObackgroundMetric}
\end{align}
Here $x$, $y$ denote the external and internal coordinates respectively, and the line elements on AdS$_6$ and $S^4$ are of unit radius.
In what follows we will use the external metric $g_{\mu\nu}$, and internal metric $g_{mn}$ defined by
\begin{align}
	\frac{9}{4} \dd s^2_{\mathrm{AdS}_6}  =  g_{\mu\nu}(x) \dd x^\mu \dd x^\nu \, , \qquad \dd s^2_{S^4}  =  g_{mn}(y) \dd y^m \dd y^n  =  \dd \theta^2 + \sin^2 \theta \dd s^2_{S^3} \, .
	\label{metrics}
\end{align}
The warp factor is 
\begin{equation}
e^{2A} = \left( \frac{3}{2} F_0 \cos \theta \right)^{1/12} \, .
\label{warp}
\end{equation}
The remaining non-zero fields of the solution are the dilaton and the 4-form field strength given by:
\begin{align}
	e^{\mathring{\Phi}}  =  \left( \frac{3}{2} F_0 \cos \theta \right)^{-5/6} \, , \qquad 
	\mathring{F}_4  =  - \frac{10}{3} \left( \frac{3}{2} F_0 \cos \theta \right)^{1/3} \vol_{S^4}  \, . \label{BObackgroundOther} 
\end{align}
The coordinate $\theta$ lies in the interval $[0,\pi/2]$, and at $\theta= \pi/2$, where the warp factor diverges, the geometry has a boundary corresponding to the location of the O8-plane. The internal space is therefore more accurately a hemisphere $HS^4$ with an
$S^3$ boundary at $\theta = \pi/2$.

\section{Linearised equations of motion}\label{sLinEOM}

In this section we consider small fluctuations around the AdS$_6 \times_w S^4$ solution outlined in the previous section, determine the equations of motion linearised in fluctuations, and reorganise them as field equations for massive free fields propagating in AdS$_6$.

We perturb the bosonic fields around their background values as:
\begin{align}
	&g_{MN} = \mathring{g}_{MN}(x,y) + e^{2A} h_{MN}(x,y) \, , \qquad \Phi = \mathring{\Phi}(y) + \phi(x,y) \, , \nn \\[5pt]
	H_3 = \mathring{0} &+ \delta H_3(x,y) \ , \qquad
	F_2 = \mathring{0} + \delta F_2(x,y) \, , \qquad 
	F_4 = \mathring{F}_4(y) + \delta F_4(x,y) \, ,
 \label{BOPert}
\end{align}
and similarly for the fermionic fields:
\begin{align}
\Psi_M = \mathring{0} + e^A \psi_M(x,y) \, , \qquad \dil = \mathring{0} + \lambda(x,y) \, .
\end{align}
The Bianchi identities \eqref{Bianchi} allow us to introduce potentials $b_2$, $a_1$ and $a_3$ such that
\begin{align}
	\qquad \delta H_3  =  \dd b_2 \, , \qquad \delta F_2  =  \dd a_1 + F_0 b_2 \, , \qquad \delta F_4  =  \dd a_3 \, , 
\end{align}
and it is in terms of these potentials that we will write the equations of motion.

In what follows all geometric quantities, in particular covariant derivatives, and contractions are with respect to the $g_{\mu\nu}$ and $g_{mn}$ metrics defined by \eqref{metrics}. In order to keep the equations covariant we will not substitute for the value of the function $A(\theta)$, given by \eqref{warp}. Also, where it occurs, we will replace the background dilaton by its equivalent value $\mathring{\Phi} = -20 A$. The Laplace--de Rham operators acting on 0-, 1-, 2- and 3-forms on AdS$_6$ are defined as
\begin{subequations}
\begin{align}
\Delta_0 \alpha &:= \nb^\mu \nb_\mu \alpha \\
\Delta_1 \alpha_\nu &:= \nb^\mu \nb_\mu \alpha_\nu - R_\nu{}^\mu \alpha_\mu \, , \\
\Delta_2 \alpha_{\nu\rho} &:= \nb^\mu \nb_\mu \alpha_{\nu\rho} - 2 R^{\mu_1}{}_{\nu}{}_{\rho}{}^{\mu_2} \alpha_{\mu_1\mu_2} + 2 R_{[\nu}{}^{\mu} \alpha_{\rho]\mu} \, , \\
\Delta_3  \alpha_{\nu\rho\sigma} &:= \nb^\mu \nb_\mu \alpha_{\nu\rho\sigma} - 6 R^{\mu_1}{}_{[\nu\rho}{}^{\mu_2} \alpha_{\sigma]\mu_1\mu_2} -3 \bR_{[\nu}{}^{\mu} \alpha_{\rho\sigma]\mu} \, ,
\end{align}
\end{subequations}
where $R_{\mu\nu} = - \frac{20}{9} g_{\mu\nu}$ and $R_{\mu\kappa\nu\lambda} = \frac{4}{9} (g_{\mu\lambda} g_{\kappa\nu} - g_{\mu\nu} g_{\kappa\lambda})$ are the Ricci and Riemann tensor of the AdS$_6$ metric $g_{\mu\nu}$. Finally, we will introduce the following notation for the warped Laplace operators on $S^4$ which appear in the linearised equations:
\begin{equation}\label{Lk}
\LL^{(k)}  := e^{-8 k A} \nb^p ( e^{8k A} \nb_p ) = \partial^2_\theta + \left(3\cot\theta - \frac{k}{3}\tan\theta\right)\partial_\theta + \frac{1}{\sin^2\theta} \Delta_{S^3} \, ,
\end{equation}
where $\Delta_{S^3}$ is the $S^3$ Laplace--Beltrami operator.

\subsection{Bosonic sector}\label{ssBosLinEOM}

\subsubsection*{Einstein equation}

We start with the Einstein equation \eqref{IIAEinstein} which splits into three subequations:
\begin{subequations}
\begin{align}
0 &= \nb^\lambda \nb_\lambda h_{\mu\nu} - 2 \nb_{(\mu} \nb^\lambda h_{\nu)\lambda} + \nb_\mu \nb_\nu h^{\lambda}{}_\lambda
+\tfrac{8}{9} h_{\mu\nu} + \LL^{(1)} h_{\mu\nu} \nn \\
&- 2 \nb_{(\mu}[e^{-8A}\nb^p(e^{8A} h_{\nu)p})] + \nb_\mu \nb_\nu h^{p}{}_p + (t^1 + t^2)g_{\mu\nu} \, , \\[10pt]
0 &= \nb^\lambda \nb_\lambda h_{\mu n} - 2 \nb_{(\mu} \nb^\lambda h_{n)\lambda} + \nb_\mu \nb_n h^{\lambda}{}_\lambda
- \tfrac{50}{9} h_{\mu n} + \LL^{(1)} h_{\mu n} \nn \\
&- 2 \nb_{(\mu}[e^{-8A}\nb^p(e^{8A} h_{n)p})] + \nb_\mu \nb_n h^{p}{}_p - 3 \cdot 2^6  A^{,p} A_{,n} h_{\mu p}  \nn \\
&- 20 A_{,n} \phi_{,\mu} - \tfrac{5}{9} e^{-8A} \epsilon_n{}^{pqr} ( \nb_\mu a_{pqr} - 3 \nb_{p} a_{qr\mu} ) \, , \\[10pt]
0 &= \nb^\lambda \nb_\lambda h_{mn} - 2 \nb_{(m} \nb^\lambda h_{n)\lambda} + \nb_m \nb_n h^{\lambda}{}_\lambda
- \tfrac{118}{9} h_{mn} + \LL^{(1)} h_{mn}  \nn \\
&- 2 \nb_{(m}[e^{-8A}\nb^p(e^{8A} h_{n)p})]  + \nb_m \nb_n h^{p}{}_p 
- 3 \cdot 2^7  A^{,p} A_{,(m} h_{n)p}  \nn \\
& - 40 A_{,(m} \phi_{,n)} + (t^1 + t^3)g_{mn} \, , 
\end{align}
\end{subequations}
where
\begin{align}
t^1  &:=   A^{,m} \left[-2 \nb^\lambda h_{m\lambda} -2  e^{-8A} \nb^n(e^{8A} h_{mn}) + \nb_m h^\lambda{}_\lambda + \nb_m h^n{}_n\right] \, , \nn \\[5pt]
t^2  &:=  \tfrac{5}{2} \left[ \tfrac{1}{24} e^{-8A} \epsilon^{mnpq} \nb_{m} a_{npq} - \tfrac{16}{45} h^{\lambda}{}_\lambda + \tfrac{17}{10} h^m{}_m  + \tfrac{96}{5} A^{,m} A^{,n} h_{mn} 
+ \left( 32 A^{,m} A_{,m} - \tfrac{7}{9} \right) \phi \right] 
\, , \nn \\[5pt]
t^3  &:=  -\tfrac{25}{6} \left[ \tfrac{1}{24} e^{-8A} \epsilon^{mnpq} \nb_{m} a_{npq} + \tfrac{7}{6} h^m{}_m  - \tfrac{288}{25} A^{,m} A^{,n} h_{mn} 
- \left( \tfrac{96}{5} A^{,m} A_{,m} - \tfrac{4}{5} \right) \phi \right] \, .
\end{align}
A comma denotes partial differentiation e.g. $A_{,m} := \partial_m A$ and $\epsilon_{mnpq}$ is the Levi--Civita tensor.

\subsubsection*{Dilaton equation}

Next we linearise the alternate form of the dilaton equation \eqref{IIADilatonAlt}:
\begin{align}
	0 &= \Delta_0 ( \phi + 2 h^\nu{}_\nu + 2 h^n{}_n ) + \LL^{(0)} ( \phi + 2 h^\nu{}_\nu + 2 h^n{}_n ) - 16 (\nabla^m \nabla^n A + 24 A^{,m} A^{,n} ) h_{mn} \nn \\
	& + 2 h_{\mu\nu} R^{\mu\nu} + 2 h_{mn} R^{mn} - 2 \nabla^\mu \nabla^\nu h_{\mu \nu} - 4 \nabla^\mu \nabla^n h_{\mu n} - 2 \nabla^m \nabla^n h_{mn} \nn \\
	& - 16 A^{,m} \big( 2 \phi_{,m} + \nabla^\nu h_{m\nu} + \nabla^n h_{mn} - \tfrac{1}{2} \nabla_m h^\nu{}_\nu - \tfrac{1}{2} \nabla_m h^n{}_n \big) \, .
\end{align}

The Ricci tensors of $g_{\mu\nu}$ and $g_{mn}$ are $R_{\mu\nu} = - \frac{20}{9} g_{\mu\nu}$ and $R_{mn} = 3 g_{mn}$.
Furthermore, the expression $\nabla_m \nabla_n A + 24 A_{,m} A_{,n}$ evaluates to
\begin{equation}
\nabla_m \nabla_n A + 24 A_{,m} A_{,n} = - \frac{1}{24} g_{mn} \, .
\end{equation}
Hence, we can recast the linearised dilaton equation in a simpler form:
\begin{align}
	0 &= \Delta_0 ( \phi + 2 h^\nu{}_\nu + 2 h^n{}_n ) + \LL^{(0)} ( \phi + 2 h^\nu{}_\nu + 2 h^n{}_n )  \nn \\
	& - \tfrac{40}{9} h^\nu{}_\nu + \tfrac{20}{3} h^n{}_n - 2 \nabla^\mu \nabla^\nu h_{\mu \nu} - 4 \nabla^\mu \nabla^n h_{\mu n} - 2 \nabla^m \nabla^n h_{mn} \nn \\
	& - 16 A^{,m} \big( 2 \phi_{,m} + \nabla^\nu h_{m\nu} + \nabla^n h_{mn} - \tfrac{1}{2} \nabla_m h^\nu{}_\nu - \tfrac{1}{2} \nabla_m h^n{}_n \big) \, .
\end{align}

\subsubsection*{2-form field strength equation}

Equation \eqref{IIA2Form} for $F_2$ splits into two subequations:
\begin{subequations}
\begin{align}
0 &= \Delta_1 a_\nu - \nb_\nu \nb^\mu a_\mu  + \LL^{(-3)} a_\nu -\nb_\nu[e^{24A} \nb^m (e^{-24A} a_m)] \nn \\ 
&+ F_0 \nb^\mu b_{\mu\nu} + F_0 e^{24A} \nb^m (e^{-24A} b_{m\nu}) \, , \\[10pt]
0 &= \Delta_0 a_n - \nb_n \nb^\mu a_\mu 
+ (\LL^{(-3)} - 3) a_n - 24 (\nb_n \nb^m A) a_m  - \nb_n[e^{24A} \nb^m (e^{-24A} a_m)] \nn \\ 
&+ F_0 \nb^\mu b_{\mu n} + F_0 e^{24A} \nb^m ( e^{-24A} b_{mn}) + \tfrac{5}{3} e^{24A} \epsilon_{n m_1m_2m_3} {\nb}^{m_1} b^{m_2m_3} \, .
\end{align}
\end{subequations}

\subsubsection*{3-form field strength equation}

Equation \eqref{IIA3Form} for $H_3$ splits into three subequations:
\begin{subequations}
\begin{align}
0 &= \Delta_2 b_{\nu\rho} + 2 \nb_{[\nu} \nb^\mu b_{\rho]\mu} + (\LL^{(3)} - F_0^2 e^{-48A}) b_{\nu\rho}
     + 2 \nb_{[\nu}[e^{-24A} \nb^m (e^{24A} b_{\rho]m})] 
\nn \\
  & - 2 F_0 e^{-48A} \nb_{[\nu} a_{\rho]} 
    -\tfrac{10}{3}\tfrac{1}{3!} e^{-16A}{\epsilon}_{\nu\rho\mu_1\dots\mu_4} \nb^{\mu_1} a^{\mu_2\mu_3\mu_4} \, , \\[10pt]
0 &= \Delta_1 b_{n \rho} + 2 \nb_{[n} \nb^\mu b_{\rho]\mu} + (\LL^{(3)} - 3 - F_0^2 e^{-48A}) b_{n\rho} - 24 (\nb_n \nb^m A) b_{\rho m} 
\nn \\
  &+ 2 \nb_{[n}[e^{-24A} \nb^m (e^{24A} b_{\rho]m})] - 2 F_0 e^{-48A} \nb_{[n} a_{\rho]} \, , \\[10pt]
0 &=\Delta_0 b_{n r} + 2 \nb_{[n} \nb^\mu b_{r]\mu} + (\LL^{(3)} - 4 - F_0^2 e^{-48A}) b_{nr} - 2 \cdot 24 (\nb_{[n} \nb^m A) b_{r]m} 
\nn \\
  &+ 2 \nb_{[n}[e^{-24A} \nb^m (e^{24A} b_{r]m})] - 2 F_0 e^{-48A} \nb_{[n} a_{r]}  
   + \tfrac{5}{3} e^{-24A} {\epsilon}_{nrm_1m_2}(2 \nb^{m_1} a^{m_2} + F_0 b^{m_1m_2}) \, .
\end{align}
\end{subequations}

\subsubsection*{4-form field strength equation}

Equation \eqref{IIA4Form} for $F_4$ splits into four subequations:
\begin{subequations}
\begin{align}
0 &= \Delta_3 a_{\nu\rho\sigma} -3 \nb_{[\nu} \nb^\mu a_{\rho\sigma]\mu} + \LL^{(-1)} a_{\nu\rho\sigma} 
    -3 \nb_{[\nu} [e^{8A} \nb^m(e^{-8A}a_{\rho\sigma]m})] \nn \\
  &- \tfrac{10}{3} \tfrac{1}{2} e^{16A} \epsilon_{\nu\rho\sigma\mu_1\mu_2\mu_3} \nb^{\mu_1} b^{\mu_2\mu_3} \, ,
    \\[10pt]
0 &= \Delta_2 a_{n\rho\sigma} -3 \nb_{[n} \nb^\mu a_{\rho\sigma]\mu} + (\LL^{(-1)} -3) a_{n\rho\sigma} 
    -8 (\nb_n \nb^m A) a_{\rho\sigma m} \nn \\
  &-3 \nb_{[n} [e^{8A} \nb^m(e^{-8A}a_{\rho\sigma]m})] \, ,
\\[10pt]
0 &= \Delta_1 a_{nr\sigma} -3 \nb_{[n} \nb^\mu a_{r\sigma]\mu} + (\LL^{(-1)} -4) a_{nr\sigma} - 2 \cdot 8 (\nb_{[n} \nb^m A) a_{r]\sigma m} \nn \\
&-3 \nb_{[n} [e^{8A} \nb^m(e^{-8A}a_{r\sigma]m})] + \tfrac{10}{3} e^{8A} {\epsilon}_{mpnr} e^{-A} \nb^{m}(e^{A} h^p{}_\sigma) \, ,
\\[10pt]
0 &= \Delta_0 a_{nrs} -3 \nb_{[n} \nb^\mu a_{rs]\mu} + (\LL^{(-1)} -3) a_{nrs} - 3 \cdot 8 (\nb_{[n} \nb^m A) a_{rs] m} \nn \\
&-3 \nb_{[n} [e^{8A} \nb^m(e^{-8A}a_{rs]m})] 
-\tfrac{5}{3}e^{8A}\phi^{,m} {\epsilon}_{mnrs} \nn \\
&+ \tfrac{10}{3} e^{8A} \left(\nb^\lambda h_\lambda{}^m + \nb^p h_p{}^m - \tfrac{1}{2} \nb^m (h^\lambda{}_\lambda+h^p{}_p) \right) {\epsilon}_{mnrs}
+ 10 e^{8A} \nb^m h^p{}_{[n} {\epsilon}_{|mp|rs]} \, .
\end{align}
\end{subequations}

\subsection{Fermionic sector}\label{ssFerLinEOM}

The equation of the gravitino \eqref{IIAGravitino} yields two subequations:
\begin{subequations}
\begin{align}
	0 &= \Gamma^{{\mu\nu_1\nu_2}} {\nabla}_{{\nu}_1} \psi_{\nu_2} + \Gamma^{{\mu}} \Gamma^{{n}_1{n}_2} \big( {\nabla}_{{n}_1} \psi_{{n}_2} + \tfrac{9}{2} A_{,n_1} \psi_{{n}_2} \big) \nn \\
	& + \Gamma^{{\mu\nu}} \Gamma^{{n}} \big({\nabla}_{{\nu}} \psi_{{n}} - {\nabla}_{{n}} \psi_{{\nu}} - \tfrac{9}{2} A_{,n} \psi_{{\nu}} \big)   - 4 A^{,n} \Gamma^{{\mu}} \psi_{{n}} 
	 \nn \\
	&+ 5 A_{,n} \Gamma^{{n}} \Gamma^{{\mu}} \lambda + \tfrac{1}{4} F_0 e^{-24A} \Gamma^{{\mu\nu}} \psi_{{\nu}} + \tfrac{1}{4} F_0 e^{-24A} \Gamma^{{\mu}} \big( \tfrac{5}{4} \lambda + \Gamma^{{n}} \psi_{{n}} \big) \nn \\	
	& - \tfrac{5}{288} \epsilon_{{n_1 n_2 n_3 n_4}} \Gamma^{{n_1 n_2 n_3 n_4}}  \big( 2 \Gamma^{{\mu\nu}} \psi_{{\nu}} + \tfrac{1}{2}  \Gamma^{{\mu}} \lambda \big)  \, , \\[10pt]
	0 &= \Gamma^{{m}} \Gamma^{{\nu_1 \nu_2}} {\nabla}_{{\nu}_1} \psi_{{\nu}_2} + \Gamma^{{m n_1 n_2}} \big( {\nabla}_{{n}_1} \psi_{{n}_2} + \tfrac{9}{2} A_{,n_1} \psi_{{n}_2} \big) \nn \\
	&+ \Gamma^{{mn}} \Gamma^{{\nu}} \big(- {\nabla}_{{\nu}} \psi_{{n}} + {\nabla}_{{n}} \psi_{{\nu}} + \tfrac{9}{2} A_{,n} \psi_{{\nu}} \big) 
	+ 4 A^{,m} \big(\Gamma^{{\nu}} \psi_{{\nu}} + \Gamma^{{n}} \psi_{{n}} \big) - 4 A^{,n} \Gamma^{{m}} \psi_{{n}} \nn \\
	&+ 5 A_{,n} \Gamma^{{n}} \Gamma^{{m}} \lambda  + \tfrac{1}{4} F_0 e^{-24 A} \Gamma^{{mn}} \psi_{{n}} + \tfrac{1}{4} F_0 e^{-24A} \Gamma^{{m}} \big( \tfrac{5}{4} \lambda + \Gamma^{{\nu}} \psi_{{\nu}} \big)\nn \\	
	&- \tfrac{5}{288} \epsilon_{{n_1 n_2 n_3 n_4}}  \Gamma^{{n_1 n_2 n_3 n_4}} \big(-2\Gamma^{mn} \psi_n + \tfrac{1}{2} \Gamma^m \lambda \big) \, ,
\end{align}
\end{subequations}
and that of the dilatino \eqref{IIADilatino} the following:
\begin{align}
	0 &= \Gamma^{{\mu}} {\nabla}_{{\mu}} \lambda + \Gamma^{{m}} {\nabla}_{{m}} \lambda
	- \tfrac{5}{288} \epsilon_{m_1 m_2 m_3 m_4} \Gamma^{m_1 m_2 m_3 m_4} 
	\big( \tfrac{3}{2} \lambda - \Gamma^{\mu} \psi_{{\mu}} + \Gamma^{{n}} \psi_{{n}} \big) \nn \\
	&+ 10 A_{,m} \Gamma^{{m}} \big( \tfrac{9}{20} \lambda - \Gamma^{{\mu}} \psi_{{\mu}} 
	 - \Gamma^{{n}} \psi_{{n}} \big) + 20 A^{,m} \psi_{{m}}  
	 - \tfrac{5}{8} F_0 e^{-24A} \big( \tfrac{21}{10} \lambda + \Gamma^{{\mu}} \psi_{{\mu}} + \Gamma^{{m}} \psi_{{m}} \big) \, . 
\end{align}
In these linearised equations the gamma matrices satisfy $\{\Gamma_\mu, \Gamma_\nu \} = 2 g_{\mu\nu}$ and $\{\Gamma_m, \Gamma_n\} = 2 g_{mn}$.

\section{Spin-2 mass spectrum}\label{sSpin2}
In this section we look at the spectrum of massive gravitons or spin-2 particles propagating in AdS$_6$. These are the transverse and traceless parts of the metric fluctuation $h_{\mu\nu}$, which we will denote by $h^{tt}_{\mu\nu}$:
\begin{equation}
\nabla^\mu h^{tt}_{\mu\nu} = 0 \, , \qquad g^{\mu\nu} h^{tt}_{\mu\nu} = 0 \, .
\end{equation}

From the linearised Einstein equation we see that it satisfies
\begin{equation}
\frac{9}{4}\nabla^\lambda \nabla_\lambda h^{tt}_{\mu\nu} 
+2 h^{tt}_{\mu\nu} + \frac{9}{4} \LL^{(1)} h^{tt}_{\mu\nu} = 0 \, ,
\end{equation}
where recall $\LL^{(1)} h^{tt}_{\mu\nu} := e^{-8A} \nabla^m\left(e^{8A}\nabla_m h^{tt}_{\mu\nu}\right)$. Taking into account the fact that the anti-de Sitter metric has radius $\frac{3}{2}$, we recognize the above equation as the equation of motion of a massive graviton of mass squared $M^2$, given by the eigenvalues of $\LL^{(1)}$: $\LL^{(1)} h^{tt}_{\mu\nu} = - \tfrac{4}{9} M^2 h^{tt}_{\mu\nu}$.

We proceed to solve the eigenvalue problem by factorizing $h^{tt}_{\mu\nu}$ as
\begin{equation}
h^{tt}_{\mu\nu}(x,y) = h^{tt}_{\mu\nu}(x)\Upsilon(y)
\end{equation}
and further expanding $\Upsilon$ in terms of $S^3$ scalar spherical harmonics:
\begin{equation}\label{factorY}
\Upsilon = \sum_{\ell=0}^\infty (\sin\theta)^\ell f_{\ell}(\theta) Y_\ell \, ,
\end{equation}
where $Y_\ell$ are $S^3$ spherical harmonics of eigenvalue $-\ell(\ell+2)$. The differential operator $\LL^{(1)}$ takes the form
\begin{equation}
\LL^{(1)} = \partial^2_\theta + \left(3\cot\theta - \frac{1}{3}\tan\theta\right)\partial_\theta + \frac{1}{\sin^2\theta} \Delta_{S^3} \, ,
\end{equation}
where $\Delta_{S^3}$ is the $S^3$ Laplace--Beltrami operator. The equation $\LL^{(1)} h^{tt}_{\mu\nu} = - \tfrac{4}{9} M^2 h^{tt}_{\mu\nu}$ thus reduces to an ordinary differential equation (ODE) for $f_\ell$:
\begin{equation}
9 \sin (2\theta) f''_\ell + 6 [9+6\ell-(10+6\ell)\sin^2\theta] f'_\ell + [4M^2 - 3\ell(3\ell+10)] \sin (2\theta) f_\ell = 0 \, ,
\end{equation} 
where a prime denotes differentiation with respect to $\theta$. We now make a change of variables to
\begin{equation}
z = \sin^2\theta \, , \qquad z \in [0,1]
\end{equation}
and the ODE becomes the hypergeometric differential equation (henceforth dropping the $\ell$ subscript):
\begin{equation}
z(1-z)\frac{d^2f}{dz^2} + [c-(a+b+1)z] \frac{df}{dz} - a b f = 0 \, ,
\end{equation}
with
\begin{equation}
a = \frac{5}{6} + \frac{\ell}{2} - \frac{1}{3}\sqrt{M^2+\left(\frac{5}{2}\right)^2} \, ,
\qquad
b = \frac{5}{6} + \frac{\ell}{2} + \frac{1}{3}\sqrt{M^2+\left(\frac{5}{2}\right)^2} \, ,
\qquad
c = 2 + \ell \, .
\end{equation}

In order to define a space of admissible solutions, we will recast the hypergeometric equation in a Sturm--Liouville form:
\begin{equation}
S f = - \lambda w(z) f \, ,
\end{equation}
where
\begin{equation}
S := \frac{d}{dz} \left(p(z) \frac{d}{dz}\right) \, , 
\qquad
p(z) := z^{\ell+2} (1-z)^{2/3} \, , 
\qquad
w(z) := z^{\ell+1} (1-z)^{-1/3} \, ,
\end{equation}
and
\begin{equation}
\lambda := \frac{1}{9}\left[M^2 - \frac{3}{2}\ell\left(\frac{3}{2}\ell+5\right)\right] \, . 
\end{equation}
We introduce the weighted inner product
\begin{equation}
 (f_1,f_2)_w := \int_0^1 f_1(z)f_2(z)w(z)dz
\end{equation}
and impose boundary conditions such that two eigenfunctions $f_1$, $f_2$ of distinct eigenvalues $\lambda_1$, $\lambda_2$ are orthogonal. We compute
\begin{align}
	(\lambda_2-\lambda_1)(f_1,f_2)_w= \int_0^1 (S f_1)f_2-f_1 (S f_2) =\left[p(f'_1 f_2-f_1 f_2')\right] \big|^1_0 \, ,
\end{align}
and hence we will impose 
\begin{equation}\label{bdrycon}
p f' f \bigl|_0 = p f' f \bigl|_1 = 0 \, . 
\end{equation}
As $p$ is zero at $z=0$ and $z=1$, the above conditions
are satisfied provided that $f$ and $f'$ are finite at the two endpoints,
or have singularities that do not dominate the zeroes of
$p$. 

Given the boundary conditions \eqref{bdrycon} we can derive a bound on the mass spectrum, as follows:
\begin{equation}
\lambda (f,f)_w = \int_0^1 (-S f) f dz = \int_0^1 p f'^2 dz \geq 0 \, ,
\end{equation}
and hence $\lambda \geq 0$ or 
\begin{equation}\label{bound}
M^2 \geq \frac{3}{2}\ell\left(\frac{3}{2}\ell+5\right) \, .
\end{equation}

Returning to the hypergeometric equation, near $z=0$, the solution is a linear combination of ${}_2F_1(a,b;c;z)$ and $(1-z)^{1-c}{}_2F_1(a-c+1,b-c+1;2-c;z)$, but since $2-c = -\ell \in \mathbb{Z}_\leq 0$, the latter needs to be replaced by a more complicated expression\footnote{See for example \cite{Olver:2010:NHM:1830479}.} which is however singular at $z=0$ and thus we discard it. We conclude:
\begin{equation}
f   = C \, {}_2F_1(a,b;c;z) \, .
\end{equation}
In order to check regularity near $z=1$ we employ the identity
\begin{equation}\label{z-1}
\begin{split}
{}_2F_1(a,b;c;z) = &\ \frac{\Gamma(c)\Gamma(c-a-b)}{\Gamma(c-a)\Gamma(c-b)} {}_2F_1(a,b;1+a+b-c;1-z) \\
+&\ (1-z)^{c-a-b} \frac{\Gamma(c)\Gamma(a+b-c)}{\Gamma(a)\Gamma(b)} {}_2F_1(c-a,c-b;1-a-b+c;1-z) \, .
\end{split}
\end{equation}
Since $c-a-b = 1/3$, we conclude that $f$ is regular at $z=1$, with
\begin{equation}
\lim_{z \to 1} f = C\,\frac{\Gamma(2+\ell)\Gamma(1/3)}{\Gamma(c-a)\Gamma(c-b)} \, .
\end{equation}
A similar check for $f' = C \, \frac{ab}{c} {}_2F_1(a+1,b+1;c+1;z)$ shows that it is singular at $z=1$:
\begin{equation}
\lim_{z \to 1} f' = C \frac{ab}{c}\,\frac{\Gamma(3+\ell)\Gamma(2/3)}{\Gamma(a+1)\Gamma(b+1)} (1-z)^{-2/3} \, .
\end{equation}
Given that $f$ and $f'$ are constant at $z=0$,  we have $pf'f\bigl|_0 = 0$. On the other hand
\begin{equation}
\lim_{z \to 1} p f' f = C^2 \frac{a b}{c} \frac{\Gamma(3+\ell)\Gamma(2/3)}{\Gamma(a+1)\Gamma(b+1)} \frac{\Gamma(2+\ell)\Gamma(1/3)}{\Gamma(c-a)\Gamma(c-b)} \, .
\end{equation}
Notice that the singularity of $f'$ at $z=1$ is of the same order as the zero of $p$, so they cancel. We would like to make 
$\lim_{z \to 1} p f' f$ vanish. There are two ways to do so:
\begin{enumerate}
\item[A.] $a = - j \in \mathbb{Z}_{\leq 0}$ in which case $f$ is a polynomial and $f'$ is regular at $z=1$. Imposing so, we derive the mass spectrum:
\begin{equation}
M^2 = \left(\frac{3}{2}\ell +3j\right)\left(\frac{3}{2}\ell + 3j +5\right) \, .
\end{equation}
$f$ becomes proportional to the Jacobi polynomial $P_j^{(\ell+1,-1/3)}(1-2z)$.\footnote{The Jacobi polynomials are defined in terms of the hypergeometric function as $P_n^{(l_1,l_2)}(x) = {l_1 + n \choose n } {}_2 F_1\left(-n,l_1+l_2+n+1;l_1+1;\tfrac{1}{2}(1-x)\right)$.}
\item[B.] $c-b = - j \in \mathbb{Z}_{\leq 0}$ in which case $\lim_{z \to 1} f = 0$ and $f'$ is singular at $z=1$. Imposing so, we derive the mass spectrum:
\begin{equation}
M^2 = \left(\frac{3}{2}\ell +3j+1\right)\left(\frac{3}{2}\ell + 3j +6\right) \, .
\end{equation}
$f$ becomes proportional to $(1-z)^{1/3} P_j^{(1/3, \ell+1)}(2z-1)$.
\end{enumerate}

We will refer to the above two branches of the mass spectrum as branch A and branch B. One might be sceptical about branch B, since in that case $f'$ is singular, but as we will see the representation theory of the $\f(4)$ superconformal algebra supports its existence. 

In particular, the Kaluza--Klein excitations should organize in multiplets of $\f(4)$, the symmetry of the AdS$_6 \times_w S^4$ solution, that include states of highest spin 2. The states lying at the bottom of the spectrum, at $j=0$, are expected to belong to short multiplets with masses determined by their $\SU(2)_R$ R-symmetry charge (spin), which for states corresponding to $S^3$ harmonics $Y_\ell$ is $\ell/2$\footnote{The R-symmetry resides in the $\SO(4)$ isometry of $S^3$.}. 

According to the AdS/CFT dictionary, the scaling dimension $\Delta$ of the operator dual to a bulk graviton excitation is given by the relation
\begin{equation}\label{mass-dim}
M^2 = \Delta(\Delta-5) \, .
\end{equation}
First note that via this relation the bound \eqref{bound} for the mass spectrum maps to a unitarity bound for the dimension of the dual field theory operators: $\Delta \geq \frac{3}{2} \ell + 5$. Furthermore, \eqref{mass-dim} gives the following dimensions for the two branches:
\begin{equation}
\text{branch A: } \ \Delta = \frac{3}{2} \ell + 3j + 5 \, , \qquad \qquad \text{branch B: } \ \Delta = \frac{3}{2} \ell + 3j + 6 \, .
\end{equation}
We thus expect short multiplets of $\f(4)$ with a state of (highest) spin 2 and dimension $\Delta = \frac{3}{2} \ell + 5$ for branch A and dimension $\Delta = \frac{3}{2} \ell + 6$ for branch B. This is indeed the case as shown in the work of \cite{Buican:2016hpb, Cordova:2016emh}:
in \cite{Buican:2016hpb} the corresponding multiplets are $\mathcal{B}[0,0;k]$ and $\mathcal{A}[0,0;k]$ in Table 1 respectively, and in \cite{Cordova:2016emh} they are $B_2$ and $A_4$ in Table 22, given explicitly in section 4.7 there.

The dual five-dimensional superconformal field theory has no Lagrangian description, but arises as the strongly-coupled UV fixed point of a Yang--Mills theory coupled to hypermultiplets. In particular the gauge group is $\USp(2N)$, and the matter content comprises $N_f < 8$ hypermultiplets in the fundamental representation and one hypermultiplet in the antisymmetric representation. We can employ the fields of this theory for a schematic description of the operators dual to the graviton modes. The operator dual to the massless graviton is of course the stress-energy tensor $T_{\mu\nu}$ with dimension $\Delta = 5$. For branch A, we can construct operators $\mathcal{O}^A_{\mu\nu}$ by multiplying $T_{\mu\nu}$ with the scalars in the hypermultiplets which transform as a doublet under SU$(2)_{\rm R}$ and have dimension $\Delta = \frac{3}{2}$. In particular we have:
\begin{equation}
\mathcal{O}^A_{\mu\nu} := {\rm Tr}\left((\epsilon^{IJ} A_I A_J)^j A_{(I_1} A_{I_2} \dots A_{I_\ell)}\right) T_{\mu\nu} \, ,
\end{equation}
where $A^I_{ab}$ is the hypermultiplet in the antisymmetric representation of $\USp(2N)$, with $I$ an SU$(2)_{\rm R}$ index and $a, b$ $\USp(2N)$ gauge indices. The trace Tr refers to the contraction of the gauge indices of $A$ which are contracted with $J_{ab}$, the $\USp(2N)$ invariant antisymmetric tensor, while $\epsilon_{IJ}$ is the SU$(2)_{\rm R}$ one.
The description of the operators dual to the graviton modes of branch B, in terms of the fields of the IR theory is less clear, if possible.
We expect these operators to have the form $\mathcal{O}^B_{\mu\nu} = \mathcal{O}^A_{\mu\nu} \mathcal{O}$, where $\mathcal{O}$ is a scalar operator of dimension one and R-charge zero. However, $\mathcal{O}$ doesn't admit a straightforward representation by the IR fields. A potential candidate for $\mathcal{O}$ would be the scalar in the vector multiplet, however the latter is not a representation of
the superconformal algebra $\f(4)$.

\section{Conclusions}\label{sConclusions}

In this note we have taken a first step towards obtaining the Kaluza--Klein mass spectrum 
of massive IIA supergravity on warped AdS$_6 \times_w S^4$. In particular, we have derived
the linearised equations of motion for fluctuations (bosonic and fermionic) around the background and determined
the mass spectrum of the spin-2 ones.  By a careful analysis of the boudary conditions of the latter at the singularity 
of the background solution, we have uncovered the existence of two branches
of mass spectra. These are bounded from below and the excitations that saturate the bound belong to short supermultiplets, 
which we have identified from the representation theory of the symmetry algebra of the solution. For one of the two branches 
we have provided an effective description of the dual field theory operators, in terms of the fields of the 
Yang--Mills--matter theory which in the strongly-coupled limit, gives rise to the  superconformal field theory. For the second branch
we lack such a description, and it would be interesting to investigate more the nature of these spin-2 
operators.

The next step in this endeavour is to determine the mass spectrum for the rest of the fluctuations.
This is a challenging task as the warped nature of the background complicates the equations of motion,
and the harmonic expansion of the modes on the internal manifold. A convenient gauge for the modes 
has to be chosen, in which the equations of motion simplify, and the form of the latter suggest
a warped generalization of the transverse gauge that is usually used for Kaluza--Klein theories
on spheres. Ultimately, we expect that the mass spectrum will be determined by the eigenvalue problem
of the warped Laplace operators $\mathcal{L}^{(k)}$, defined in \eqref{Lk}. As was the case for the spin-2 modes, for which
$k=1$, this eigenvalue problem can be mapped to a hypergeometric differential equation; see appendix \ref{Lkspec}.

\section*{Acknowledgements}
We would like to thank Joseph Hayling, Noppadol Mekareeya,  Costis Papageorgakis, Diego Rodr\'{i}guez-­G\'{o}mez, Alessandro Tomasiello and Alberto Zaffaroni for useful correspondence and discussions. AP is supported by the Knut and Alice Wallenberg Foundation under grant Dnr KAW 2015.0083. PR was partially supported by the INFN.

\appendix

\section{Identities}

\subsection{Metric perturbations}

Under a small perturbation of the metric $g_{MN} \rightarrow g_{MN} = \mathring{g}_{MN} + \delta g_{MN}$, the inverse metric transforms as $g^{MN} = \mathring{g}^{MN} - \delta g^{MN}$. The Christoffel symbols $\Gamma^P_{MN}$, Laplace operator $\nb^2 : = \nb^M \nb_M$, and Ricci tensor $R_{MN}$ transform as:
\begin{align}
	\Gamma^P_{MN} \ =& \ \mathring{\Gamma}^P_{MN} + \tfrac{1}{2} \left( \mathring{\nabla}_M \delta g_N{}^P + \mathring{\nabla}_N \delta g_M{}^P - \mathring{\nabla}^P \delta g_{MN} \right) \, , \\
	\nb^2 \ =& \ \mathring{\nb}^2 - \delta g^{MN} \mathring{\nabla}_M \partial_N - \left( \mathring{\nabla}^M \delta g_M{}^N - \tfrac{1}{2} \mathring{\nabla}^N \delta g_M{}^M \right) \partial_N \, , \label{WarpLap} \\
	R_{MN} \ =& \ \mathring{R}_{MN} + \tfrac{1}{2} \mathring{\Delta}_L \delta g_{MN} + \mathring{\nabla}_{(M} \mathring{\nabla}^P \delta g_{N)P} - \tfrac{1}{2} \mathring{\nabla}_M \mathring{\nabla}_N \delta g_P{}^P \, , \label{WarpRicci}
\end{align}
where the Lichnerowicz operator is defined as
\begin{align}
	\mathring{\Delta}_L \delta g_{MN} \ :=& \ - \mathring{\nb}^2 \delta g_{MN} - 2 \mathring{R}_{MPNQ} \delta g^{PQ} + 2 \mathring{R}_{(M}{}^P \delta g_{N)P} \, , \label{WarpLich}
\end{align}
and all indices are raised and lowered using the metric $\mathring{g}$.

\subsection{Conformal transformations}

The Christoffel symbols $\mathring{\Gamma}^P_{MN}$ associated with the warped background metric $\mathring{g}_{MN}$ can be expressed in terms of the Christoffel symbols $\Gamma^P_{MN}$ associated with the unwarped metric $g_{MN} = e^{-2A} \mathring{g}_{MN}$ and derivatives of the warp factor $A$ as follows:\footnote{A comma denotes a partial derivative: $A_{,M} := \partial_M A$.}
\begin{align}
	\mathring{\Gamma}^P_{MN} = \Gamma^P_{MN} + \delta^P_M A_{,N} + \delta^P_N A_{,M} - g_{MN} A^{,P} \, .
\end{align}
Indices on the right-hand side are raised using $g_{MN}$. For the Riemann and Ricci tensors the analogous expressions are:
\begin{align}
\mathring{R}^M{}_{NPQ} &=  R^M{}_{NPQ} + 2 \delta^M_{[Q}\nb_{P]} A_{,N} + 2g_{N[P} \nb_{Q]} A^{,M} \nn \\
& \hspace{1cm} + 2 \delta^M_{[P} A_{,Q]} A_{,N} + 2 g_{N[Q} A_{,P]} A^{,M} + 2 \delta^M_{[Q} g_{P]N} A_{,R} A^{,R} \, , \\
\mathring{R}_{MN} &= R_{MN} + 8(A_{,M} A_{,N} - \nb_M \nb_N A) - (8 A_{,P} A^{,P} + \nb^2 A) g_{MN} \, .
\end{align}

The relations between warped and unwarped Laplace operators and covariant spinor derivatives are:
\begin{align}
	\mathring{\nb}^2  &=  e^{-2A} \left(\nb^2 + 8 A^{,M} \partial_M \right) \, , \\
	\mathring{\nabla}_M &= \nabla_M + \frac{1}{2} A_{,N} \Gamma_M{}^N \ ,
\end{align}
where the gamma matrices $\Gamma_M$ satisfy $\{ \Gamma_M, \Gamma_N \} = 2 g_{MN}$.

A warped covariant derivative of the metric perturbation $\delta g_{MN} = e^{2A} h_{MN}$ decomposes as
\begin{align}
	e^{-2A} \mathring{\nabla}_Q \delta g_{MN}  :=   T_{QMN}  =  \nb_Q h_{MN} - 2 A_{,(M} h_{N)Q} + 2 A^{,R} g_{Q(M} h_{N)R} \, .
\end{align}
Using the above tensor $T$ we find:
\begin{align}
	e^{-2A} \mathring{\nabla}_P \mathring{\nabla}_Q \delta g_{MN}  &=  \nabla_P T_{QMN} - 2 A_{,(P} T_{Q)MN} - A_{,M} T_{QPN} - A_{,N} T_{QPM} \nn \\
	&+ g^{RS} A_{,S} (g_{PQ} T_{RMN} + g_{PM} T_{QRN} + g_{PN} T_{QRM} )  \, .
\end{align}
Employing these results we can derive expressions for all warped quantities in \eqref{WarpLap}, \eqref{WarpRicci} and \eqref{WarpLich} in terms of unwarped ones.

Finally, we record the ``unwarping'' of the following terms that appear in the equations of motion of the fermions:
\begin{align}
	\mathring{\Gamma}^M \mathring{\nabla}_M \Lambda &= e^{-A} \big( \Gamma^M \nabla_M \Lambda + \tfrac{9}{2} A_{,M} \Gamma^M \Lambda \big) \, , \\
	\mathring{\Gamma}^{MNP} \mathring{\nabla}_{N} \Psi_{P} &= e^{-3A} \big( \Gamma^{MNP} \nabla_{N} \Psi_{P} + \tfrac{7}{2} A_{,N} \Gamma^{MNP} \Psi_{P} + 8 A^{,[M} \Gamma^{N]} \Psi_N \big) \, ,
\end{align}
where on the left-hand side the gamma matrices $\mathring{\Gamma}_M$ satisfy $\{ \mathring{\Gamma}_M , \mathring{\Gamma}_N \} = 2 \mathring{g}_{MN}$.

\section{The operators $\mathcal{L}^{(k)}$}\label{Lkspec}
The eigenvalue equation
\begin{equation}
\LL^{(k)} \varphi = - \frac{4}{9} M^2 \varphi \, ,
\end{equation}
for the operators $\LL^{(k)}$ defined in \eqref{Lk}, upon introducing $f(\theta)$ such that $\varphi = (\sin\theta)^\ell f$ and switching variables to
 \begin{equation}
z = \sin^2\theta \, , \qquad z \in [0,1] \, ,
\end{equation}
becomes the hypergeometric differential equation
\begin{equation}
z(1-z)\frac{d^2f}{dz^2} + [c-(a+b+1)z] \frac{df}{dz} - a b f = 0 \, ,
\end{equation}
where $c = 2 + \ell$ and 
\begin{equation}
a = \frac{9+k}{12} + \frac{\ell}{2} +\frac{1}{3}\sqrt{M^2+\left(\frac{9+k}{4}\right)^2} \,  ,
\qquad
b = \frac{9+k}{12} + \frac{\ell}{2} - \frac{1}{3}\sqrt{M^2+\left(\frac{9+k}{4}\right)^2} \,   .
\end{equation}

\newpage

\bibliographystyle{utphys}
\bibliography{6dKKbib}

\end{document}